# Strain dependent elastic modulus of graphene

*Guillermo López-Polín*[†1], Miriam *Jaafar*[†2], *Francisco Guinea*[3,4], *Rafael Roldán*[2]
*Cristina Gómez-Navarro*[1,5]*and Julio Gómez-Herrero*[1,5]

[1] Departamento de Física de la Materia Condensada, Universidad Autónoma de Madrid, 28049, Madrid, Spain.

[2] Instituto de Ciencia de Materiales, CSIC, 28049, Madrid, Spain.

[3] Department of Physics, University of Manchester, M13 9PL, Manchester, UK.

[4] Instituto IMDEA Nanociencia, 28049, Madrid, Spain.

[5] Centro de Investigación de Física de la Materia Condensada, Universidad Autónoma de Madrid, 28049, Madrid.

[†]Both authors contributed equally.

*Correspondence to: cristina.gomez@uam.es

**Indentation experiments on graphene membranes pre-stressed by hydrostatic pressure show an increase in effective elastic modulus from ∼*300 N/m* in non pressurized membranes to ∼*700 N/m* for pre-strains above *0.5 %*. This pronounced dependence of the stiffness of graphene with strain is attributed to its high anharmonicity and the great influence of out of plane corrugations of this atomic thick membrane in its mechanical properties. Our experimental findings imply that graphene´s measured stiffness is highly influenced by the presence of corrugations and that the in plane elastic modulus corresponding to atomic bond stretching is more akin to *700 N/m*, instead of the commonly accepted *340 N/m*.**[1-5]

Supplementary information downloadable at:

http://intranet.fmc.uam.es/supplementary_info/Press_memb_SI_def.pdf

Membranes are the paradigm of two-dimensional materials. Mimicking other low dimensional materials, such as 0D quantum dots and 1D wires, reduced dimensionality in thin membranes entails new emergent phenomena. Bendable membranes (with bending rigidity comparable to their thermal energy) exhibit entropic effects in the form of out of plane fluctuations (OPF) that bring out exotic mechanical properties[6]. The elastic theory of membranes has a long history beginning with Poisson work in the early 19$^{th}$ century, followed by Landau and Lifshitz advances and a great boost given to these models during the 70s due the high interest in biological tissues.[6-9] One of the most known consequences of the presence of these OPF is the negative thermal expansion coefficient exhibited by these membranes. An even more exotic property of such membranes is the renormalization of their elastic constants[10]; according to these models the elastic modulus of membranes are highly influenced by the energy required to "flatten out" these fluctuation and not only dictated by atomic bond stretching as in conventional materials. Indeed, due to the strong anharmonic interaction between bending and stretching modes, which dominates the elastic properties of 2D membranes at long wavelengths, stiffness should tend to zero when the size of the membrane tends to infinite and it should increase with increasing strain.[6, 11]

This unusual behavior of elastic constants can be visualized considering the static counterpart of OPF: wrinkles. The force needed to stretch a macroscopic crumpled sheet is smaller than that required to stretch a flat one. The elastic modulus of the flat sheet is usually called bare elastic modulus, while the one corresponding to the wrinkled sheet is the renormalized one. For the bending rigidity it is just the opposite; if the flat sheet is suspended by one corner it bends down much easier than the crumpled sheet. The importance of thermal OPF is nicely captured by a magnitude related to the Föppl-von Karman number $\gamma_T = (E_{2D} k_B T \ell^2)/\kappa^2$ being $E_{2D}$ the in plane elastic modulus, $\ell$ a characteristic length and $\kappa$ the bending rigidity. Large values of $\gamma_T$ implies strong renormalization of the elastic constants.

Graphene is the nature's thinnest elastic membrane. It is highly bendable, ($\kappa \sim 1$ eV, [1, 2, 12, 13,11]), stiff and anharmonic. Therefore the above mentioned phenomena should apply to it[14]. This is the case of the negative thermal expansion coefficient, which has already been measured[15, 16] and addressed theoretically [17]. Graphene $\gamma_T$ is already much greater than 1 for $\ell > 1 nm$ therefore strong renormalization of its elastic constants is expected at room temperature. So far, these effects have rarely been observed; elastic measurements performed in different groups on graphene membranes of sizes ranging from 0.5 up to 5 μm have not shown size dependent elastic constants. External induced pre-strains as substantial as 0.3% [2, 18, 19] revealed no dependence of the measured elastic constants. A recent work by López-Polín *et al.* [20] reported an increment on the EM of suspended graphene of almost a factor of two when a density of vacancies of ~0.2% is induced. This result suggests that the elastic modulus of graphene is dressed by the presence of OPF; upon introduction of a new characteristic length in the membrane, that is the average distance between vacancies of ~ 4 nm, longer wavelength oscillations are diminished unraveling the non-renomralized elastic modulus of graphene.

Here we show than when graphene membranes are pre-stressed by hydrostatic pressure, the elastic modulus presents an increasing tendency with strain. Above a certain stress this figure saturates at a value about 700 N/m, that we consider as the bare (non-renormalized) elastic modulus of graphene.

In order to investigate the importance of the above commented effects suspended membranes were prepared by mechanical exfoliation of natural graphite on $SiO_2$(300nm)/ Si substrates with predefined circular wells with diameters ranging from 0.5 to 3 µm (see SI1). Single layers of graphite were found by optical microscopy and corroborated by Raman spectroscopy [21] (see SI2). Raman spectra also confirmed absence of defects in the suspended membranes. Non contact Atomic Force Microscopy (AFM) images of the drumheads revealed that graphene layers adhere to the vertical walls of the wells for 2-10 nm in depth in their initial morphology (See Figure 1). For these experiments we selected membranes showing a flat and featureless topography with no initial noticeable slack. Indentations with AFM tips (see SI3) were performed on the center of the circular membranes. In order to avoid membrane breaking the maximum applied force was kept below 1200nN. Repeated loading/unloading cycles on the same membrane showed high reproducibility, completely reversible behavior and no signature of fatigue. The resulting Force *vs.* Indentation ($F(\delta)$) curves allow determining the effective elastic modulus of graphene ($E_{2D}$). Indentation curves showed good fitting to the Schwerin equation, commonly applied to point loaded clamped circular membranes[2, 22]

$$F(\delta) = \pi \sigma_0 \delta + \frac{E_{2D}}{a^2} \delta^3 \qquad \text{eq 1}$$

where $F$ is the loading force, $\delta$ is the indentation at the central point, $\sigma_0$ is the pretension accumulated in the sheet during the preparation procedure and $a$ is the well radius. In the absence of hydrostatic pressure the obtained values of $E_{2D}$ ranged between *250* and *360 N/m* (see SI2) and $\sigma_0$ from 0.05-0.6 N/m (pre-strain 0.01-0.15%) with no correlation between pre-stress and measured $E_{2D}$.

These drumheads have been previously shown to behave as well sealed micro-cavities due to the high adhesion of graphene to the $SiO_2$ substrate.[18, 23] Global stress in the suspended graphene can be achieved by applying a pressure difference across the membrane. To this end, samples were inserted in a chamber (base pressure $10^{-6}$ mbar) equipped with an atomic force microscope. The pressure in this chamber can be varied from ~0 to 4 atm. Samples were kept in this chamber at $10^{-6}$ mbar for 12 hours until the pressure inside the micro cavity, defined by the $SiO_2$ and the graphene layer, equilibrates to the external chamber pressure. Then, an overpressure of $P_{out}$ =3 atm of $N_2$ was set in the chamber; in this manner we obtained a total pressure difference across the membrane ($\Delta P= P_{out}-P_{in}$) of about 4 atm. As we shall show below, this $\Delta P$ induces externally induced strains up to 0.6% in our graphene blisters. Figure 1 illustrates different shapes of a representative membrane for various $\Delta P$. The graphene membranes show concave or convex geometry depending on the sign of $\Delta P$. In order to track both the strain and elastic modulus, upon introduction of the $N_2$ overpressure, we image and indent the membranes with an AFM tip as the pressure inside and outside the micro-cavities equilibrates (see SI4).

The strain (in %) induced by pressure can be estimated from the AFM images of the membranes under pressure as.

$$\varepsilon_p = (s-2a)/(2a)\cdot 100 = [(a^2+\delta_0^2)/\delta_0 \arcsin(2\delta_0 \cdot a/(a^2+\delta_0^2))-2a]/(2a)\cdot 100 \qquad \text{eq 2}$$

Where $\varepsilon_p$ is the strain as determined by geometry, $s$ the length of a great arc of the pressurized membrane, $a$ the radius of the hole and $\delta_0$ the distance from surface plane to the center of the membrane induced by $\Delta P$ *(see SI5)*.

Analysis of indentation curves on pressurized membranes is more complex than non-pressurized ones since there is no closed-form analytical solution that accounts for this situation.[2] In order to address this issue, we performed finite element numerical calculations to simulate nano-indentations on membranes under pressure. We used *Comsol Multiphysics 4.3*, modeling a circular clamped membrane under hydrostatic pressure (see SI5). The $F(\delta)$ curves obtained by the finite element analysis with a predefined $E_{2D}$ of 340 N/m showed good fit to a full 3$^{rd}$ order polynomial expression

$$F(\delta) = c_0 + c_1 \cdot \delta + c_2 \delta^2 + c_3 \delta^3 \qquad \text{eq 3}$$

where, as in expression (1), the effective elastic modulus can be extracted from the third polynomial coefficient as $c_3 = E_{2D}/a^2$. Therefore we conclude that the effective elastic modulus of pressurized membranes can be inferred from indentation curves (see SI5). Similar fittings have been successfully used in other reports even for $\Delta P=0$.[19]

Figure 2a is a schematic drawing of our experiments in pressurized membranes. Note that indentations are always performed for $P_{out} \geq P_{in}$ ($\Delta P>0$). Figure 2b presents three representative $F(\delta)$ curves and fittings to expression 3. The obtained values of $E_{2D}$ according to the above described fittings as a function of external strain are displayed in Figure 3a. Here we can observe that for pressure induced strains ($\varepsilon_p$) lower than *0.2-0.3 %* the effective stiffness of graphene membranes does not depend on strain, in good agreement with previous results.[1, 2, 24] Above these $\varepsilon_p$ the effective elastic modulus increases with increasing $\varepsilon_p$ until a maximum value of $E_{2D} \sim 700 N/m$ where this figure shows a saturation tendency.

The above described results indicate that graphene´s elastic modulus is ill-defined for our intermediate range of $\varepsilon_p$. As indentations themselves induce strain, the measured effective elastic modulus also exhibits this dependence with indentation force in this pressure range. Figure 3 portrays this dependence; it displays the $E_{2D}$ obtained from fitting our indentation curves to equation 3 up to different indentation forces. Here we can clearly appreciate that while the values obtained at our higher and lower $\varepsilon_p$ are very robust, for the case of intermediate $\varepsilon_p$ the obtained value of the effective elastic modulus of graphene depends on the maximum indentation force used to perform the fitting. Higher indentation range yields to larger values of $E_{2D}$. This analysis further supports the notion of an ill-defined elastic modulus and corroborates that not only the strain induced by the pressure difference ($\varepsilon_p$) but also that introduced during the indentation experiments alter the elastic response of the membrane. While the effective elastic modulus of graphene cannot be properly defined in our intermediate range of $\varepsilon_p$, in order to assign a representative number to our measurements, the values of $E_{2D}$ plotted on Figure 2b have been all estimated for maximum tip load of 1000nN.

An interesting issue to tackle is weather the different kinds of induced strain equally affect the elastic modulus of graphene membranes. The ideal measurements for our purpose would involve the use of "graphene elastometer". Unfortunately they have not been fully developed yet and inducing pure uniaxial strain in one atom thick membranes still remains an issue. Approaches like the ones used in this work have been used in literature in order to circumvent this problem; however they involve different sources of strains. In our case three major kind of strains are present; pre-strain accumulated during sample preparation, global strain achieved by pressure difference and the strain induced by indentations, which is not uniform in all the membranes. None of these strains is purely uni- or bi-axial and their relative influence in quenching oscillations might be quite different. Indeed, our results suggests that $\varepsilon_p$ is more effective in increasing elastic modulus than those introduced along indentation curves or caused during sample preparation. This explains why these effects have not been observed in previous reports.[2]

The experimental results described above are along the line suggested by Roldán *et al.* in ref. [11] where a combination of atomistic Monte Carlo simulations for a graphene layer under tension, and elastic theory of membranes within the self-consistent screening approximation[10] showed that strain suppresses anharmonic coupling between bending and stretching modes. This suppression changes the dispersion of the flexural modes, shifting them to higher frequencies and quenching the lower frequency modes. As a result, there is an enhancement of the stiffness of graphene, with the corresponding increase of the in plane elastic constants. In the framework of this theory, and due to the effect of anharmonicities, the elastic constants become scale dependent. Therefore the $E_{2D}$ of graphene (which will be wave-vector dependent if we are in the anharmonic regime) can be expressed as[25]

$$E_{2D}(k) = \frac{E_{2D}^0}{1 + E_{2D}^0 \cdot b(k)}$$

where $E_{2D}^0$ is the elastic modulus at T=0 (non-renormalised) and $k$ is the momentum of the out-of-plane modes, and the function $b(k)$ accounts for the anharmonic interaction between in-plane and flexural modes, which can be calculated from the correlation functions of the flexural modes (see I7). In the absence of strain and for long wavelengths, the Young modulus is suppressed as $E_{2D}(k)$~$k^{2-2\eta}$, where $\eta$ is a characteristic exponent which, within the self-consistent screening approximation, takes the value $\eta \approx 0.82$[10]. Importantly, in the absence of any other scale that cuts off the fluctuations at long wavelengths, the theory predicts a vanishing of the elastic modulus in the *k→0* limit. However realistic 2D membranes always have a length scale that cuts off this behavior, such as the finite size of the sample. The above reduction of the stiffness can be reversed by increasing the external strain applied to the membrane which suppresses, above some characteristic wavelength $\ell_s$, the effect of anharmonicities in the membrane,[11] *driving* the elastic modulus towards its harmonic limit $E_{2D}^0$, in agreement with the experimental results shown in Fig. 3. This idea is sketched in Figure 4.

According to this theoretical framework, we interpret our experimental results as follows. Small external strains are able *only* to suppress fluctuations of very long wavelengths, longer than the characteristic graphene wavelength $\ell$ that imposes the above mentioned infrared momentum cut-off $k_{min} \sim 1/\ell$. where $\ell$ is a characteristic length that depends on the graphene peculiarities, such as for wavelengths larger than this number the EM is independent of the wavelength of the OPF. This cut off is not predicted by the elastic theory of membranes and might be only due to graphene peculiarities. This notion of a minimum relevant wavelength agrees with the experimental observation of ref.[20] in which vacancies created in graphene with an average separation between 20 and 5 nm were shown to reduce the entropic effects of the graphene membrane. We interpret this cutoff as associated to the existence of intrinsic ripples, which are of the order of tens of *nm*, as measured by TEM[26] and are also likely to influence the elastic modulus. Therefore, the E$_{2D}$ measured for small strains is the one associated to $k_{min}$, $E_{2D}(k_{min}) \sim 300\ N/m$. As we increase $\varepsilon_p$, we will overcome this cutoff and fluctuations of wavelength shorter than $\ell$ will start to be suppressed, stiffening the membrane and enhancing the E$_{2D}$. This corresponds to the intermediate range in Fig. 3a, in which the measured value of *E$_{2D}$* grows with the applied strain. This enhancement of *E$_{2D}$* continues until we reach the *harmonic* regime of the membrane, for which all the anharmonicities have been suppressed and the measured EM is the *bare* one, regardless the applied strain. This is seen by the plateau at *~700 N/m* that we interpret as the *bare* elastic modulus of graphene.

Notice that the existence of this infrared cutoff would also explain why none of the experiments performed so far in suspended membranes have reported a dependence of the elastic modulus of suspended graphene with sample size: all the experiments are carried out in membranes with characteristic lengths above 500 nm, larger than the relevant wavelength scales suggested by our experiments. While this theoretical framework enlightens the experimental findings, the calculations performed so far are based on a perturbative approach and additional work is needed for good quantitative agreement of theoretical and experimental results. (see SI7 for detailed explanation).

While our results are qualitatively well described by the elastic theory of membranes, they leave some open questions. One of the main issues to be addressed is the discrepancy between our measured value of "zero temperature" (non-renormalized) elastic modulus of graphene (~700N/m) and that calculated by first principle calculations such as Density Functional Theory, which is between 300 and 400 N/m. [5] While for conventional solids DFT yield elastic moduli with very good accuracy (usually well below 10%) our results suggest that for 2D materials it might require some fine tune [27].

The impact of these out of plane fluctuations is not limited to the mechanical properties of graphene; properties such as thermal conductivity and thermal expansion should be dominated by these flexural modes. Also charge carriers are expected to be influenced by their presence[28, 29], and comprehensive understanding of nano-electromechanical devices[30] will require a detailed knowledge of the anharmonic properties of graphene sheets. Indeed the control of graphene deformations can lead to novel ways of modifying the properties of graphene.

In summary we provide indentation measurements on graphene pressurized membranes. The elastic modulus in these membranes is not a constant magnitude but a strain dependent figure and cannot be defined as in conventional materials. This experimental observation is understood in the framework of the elastic theory of membranes where such effects have been predicted for thin bendable membranes due to high impact of out of plane fluctuations in the mechanical properties of such entities, making the elastic constants to be scale dependent. Our work supports a zero temperature (non-renormalized) elastic modulus of ~700 N/m, twice that calculated by first principle calculations[2, 4, 5], and suggests that numbers have to be revised. Due to the large effects observed, comprehensive and quantitative understanding of our results would imply non-perturbative approaches of self-screening approximation or atomistic calculation with graphene slabs more than 20 nm in lateral size. The effect observed here should also be expandable to other 2D materials, therefore experiments on different type of membranes would certainly help to clarify this topic.

**ACKNOWLEDGMENT**

We acknowledge fruitful discussions with M. Katsnelson, A. Fasolino and P. San José and S. Marquez an R. P. Pelaez for technical support. Financial support was received from P2013/MIT-3007, MAT2013-46753-C2-2-P, Consolider CSD2010-0024 and the European Research Council Advanced Grant, #290846. RR acknowledges financial support from Juan de la Cierva program (MINECO, Spain).

**FIGURE 1**

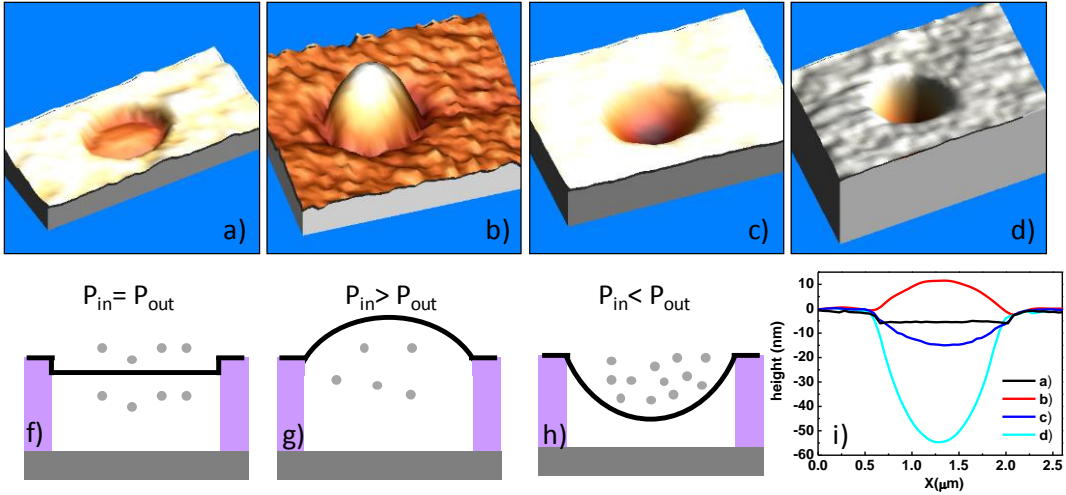

*Figure 1: **Morphology of pressurized membranes**. Panels a to d display AFM images of a graphene drumhead of 1.5 μm diameter subjected to different ΔP. a) corresponds to $P_{in}= P_{out,}=1at$. b) to $P_{out}\sim 10^{-5}$, $P_{in,}=1atm$, c) to $P_{in}\sim 10^{-5}$, $P_{out,}=1atm$ of $N_2$ and d) to $P_{in}\sim 10^{-5}$, $P_{out,}=3\ atm$ of $N_2.$ Panel f to h are schematic profiles of the pressurized membrane. Panel i) shows AFM profiles taken along a great arc in the AFM images shown in panel a to d. In this work only membrane with the configuration shows in a) and c are indented.*

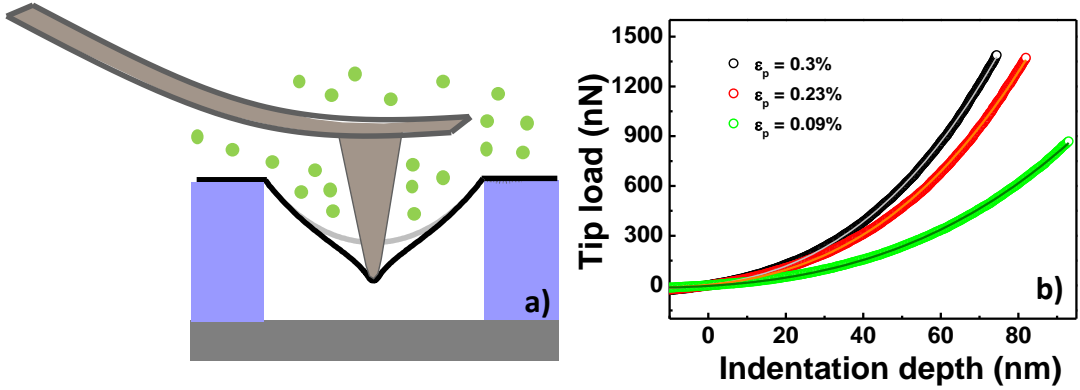

*Figure 2. **Indentations on pressurized membranes***. *a) Sketch of the indentation geometry on a pre-strained membrane. b) Empty circles are the experimental F(δ) curves on a drumhead subjected to different ΔP. Strains induced by pressure are calculated according to expression 2 in the main text. The lines are fits to the experimental data following eq. 3*

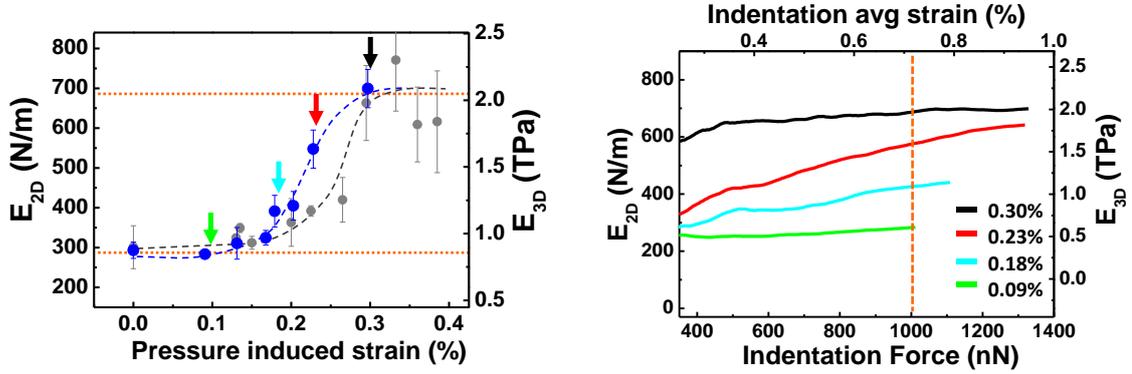

*Figure 3. Elastic modulus as a function of induced strain.* a) $E_{2D}$ as a function of induced strain by pressure difference across the membranes. Values of the 2D elastic modulus have all been obtained from fitting of the indentation curves for maximum tip load of 1000nN. Tip loads lower than 350 nN do not yield stable fittings. The dashed lines are drawn to guide the eye to the two values where we observe elastic modulus not dependent of strain, at our lowest and highest induced strains. The 3-dimensional elastic modulus has been calculated as $E_{2D}/0.34$ nm. b) Vertical axis corresponds to the value of $E_{2D}$ obtained by fitting the experimental $F(\delta)$ curves up to different indentations forces (horizontal axis). The indentation average strain (upper horizontal axis) has been calculated as $\frac{\sqrt{a^2+\delta^2}-a}{a} \cdot 100$. The four curves show four representative data set obtained on the same drumhead under different strains induced by pressure. The dashed line represents the force used to calculate the $E_{2D}$ represented in panel a). Each line corresponds to a single point in panel a). See color correspondence with the arrows.

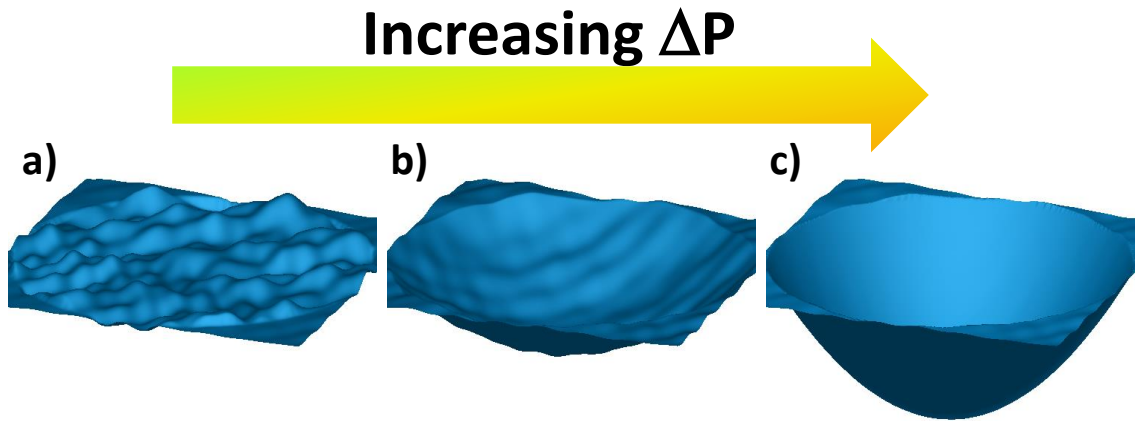

**Figure 4.** *Flexural modes in membranes suppressed by pressure induced strain.. a) Schematic view of a suspended membrane on a circular hole with ΔP=0. b) and c) reflect how wrinkles are flatten out as ΔP increases.*